\documentclass{acm_proc_article-sp}
\usepackage{times}
\usepackage[latin1]{inputenc}
\usepackage{amsmath}
\usepackage{amsfonts}
\usepackage{amssymb}
\usepackage{verbatim}
\usepackage{algorithm}
\usepackage{algorithmic}
\usepackage{graphicx}
\usepackage{hyperref}

\begin{document}
\title{Large-Scale Continuous Subgraph Queries on Streams}
\numberofauthors{4} 
\author{
%
%
\alignauthor
Sutanay Choudhury\\
       \affaddr{Pacific Northwest National Laboratory}\\
       \affaddr{902 Battelle Boulevard}\\
       \affaddr{Richland, WA, USA 99352}\\
       \email{sutanay.choudhury@pnnl.gov}
\alignauthor
Lawrence Holder\\
       \affaddr{School of Electrical Engineering \& Computer Science}\\
       \affaddr{Washington State University}\\
       \affaddr{Pullman, WA, USA 99164-2752}\\
       \email{holder@wsu.edu}
\alignauthor
George Chin\\
       \affaddr{Pacific Northwest National Laboratory}\\
       \affaddr{902 Battelle Boulevard}\\
       \affaddr{Richland, WA, USA 99352}\\
       \email{george.chin@pnnl.gov}
\and  
\alignauthor
John Feo\\
       \affaddr{Pacific Northwest National Laboratory}\\
       \affaddr{902 Battelle Boulevard}\\
       \affaddr{Richland, WA, USA 99352}\\
       \email{john.feo@pnnl.gov}
}
\maketitle
\thispagestyle{empty}
\begin{abstract}
Graph pattern matching involves finding exact or approximate matches for a query subgraph in a larger graph.  It has been studied extensively and has strong applications in domains such as computer vision, computational biology, social networks, security and finance.  The problem of exact graph pattern matching is often described in terms of subgraph isomorphism which is NP-complete.  The exponential growth in streaming data from online social networks, news and video streams and the continual need for situational awareness motivates a solution for finding patterns in streaming updates.  This is also the prime driver for the real-time analytics market.  Development of incremental algorithms for graph pattern matching on streaming inputs to a continually evolving graph is a nascent area of research.  Some of the challenges associated with this problem are the same as found in continuous query (CQ) evaluation on streaming databases.  This paper reviews some of the representative work from the exhaustively researched field of CQ systems and identifies important semantics, constraints and architectural features that are also appropriate for HPC systems performing real-time graph analytics.  For each of these features we present a brief discussion of the challenge encountered in the database realm, the approach to the solution and state their relevance in a high-performance, streaming graph processing framework.

\end{abstract}
\section{Introduction}
Graph pattern matching is defined as the problem of searching a graph for all instances of a pattern that is also expressed as a graph \cite{conte2004thirty}. This is mathematically defined as the problem of subgraph isomorphism, which given a  pattern or query graph  $G_q$ and a larger input graph $G_d$, requires finding all isomorphisms of $G_q$ in $G_d$.  Following the definition of isomorphism, the matching involves finding a one-to-one correspondence between the vertices of a subgraph of $G_d$ and vertices of $G_q$ such that all vertex adjacencies are preserved.  Dynamic graphs refer to graphs that evolve over time through addition or deletion of vertices and edges.   Therefore, the problem of graph pattern matching for dynamic graphs can be described as the \textit{continuous} process of searching for patterns in the graph as it is updated.  News \cite{Zhao:2007:TIF:1619797.1619886}, finance \cite{Chandramouli:2010:HDP:1920841.1920873}, cyber security and intelligence \cite{Coffman:2004:GTI:971617.971643} are among the primary domains that drive the real-time analytics market \cite{twitter_storm, Peng:2010:LIP:1924943.1924961} and motivate development of HPC systems.  These domains present data sources that lend themselves naturally to a graph based representation and additionally, provide semantic information in terms of types, labels and timestamps, which can be more generally described as attributes of the vertices and edges of the graph.  The availability of the attributes influence the isomorphism computation because assigning a correspondence between a pair of vertices in the search and query graph requires them to satisfy equality constraints on type and possibly, other attributes as well.  All these domains are also characterized by massive streaming data that are continuously providing updates from social networks, financial markets and malicious activities on the internet with a high emphasis on time-to-insight, the capability of learning about an event as soon as it happens.  This motivates our investigation of subgraph pattern matching on streams using  high-performance computing architectures.  
\subsection{Continuous Queries in Databases}
This problem can also be described in general as computing a function $f$ over a stream $S$ over time and notifying the user whenever the output of $f$ satisfies a user-defined constraint.  A \textit{continuous query system} is defined as one where a query logically runs continuously over time as opposed to being executed intermittently and then running to completion \cite{Terry:1992:CQO:141484.130333}.  There are some obvious challenges associated with computing continuous queries \cite{Babcock:2002:MID:543613.543615}.  In the following, we state them as two focus areas from our perspective.  

 \textbf{Area I} The memory requirement for computing $f$ and maintaining its partial results over time may be unbounded.  Adding constraints, such as restricting the class of queries to control their complexity or using approximation techniques, is often the solution to this problem.  Incorporation of temporal aspects such as specifying order of arrival for entities in the data are examples of such efforts to reduce memory requirements.  We view this as an area focusing on identifying practical constraints found in real-life applications to make the problem tractable. 

\textbf{Area II} Timeliness requirements of query processing is set by the underlying application or the end-user.  Typically there exists a data source (e.g. a relational or graph database) that needs to be updated with the streaming input and computation of $f$.  The relative cost of updating the data source and computation of $f$ strongly influences processing strategies such as on-the-fly processing, batch processing, incremental processing, sampling etc. We broadly view this under the category of algorithmic challenges.
Continuous queries are also distinguished from ad-hoc query processing by their high selectivity (looking for unique events) and need to detect newer updates of interest as opposed to retrieving lots of past information.  Conventional databases are passive repositories with large collections of data that works in a request-response model, whereas continuously monitoring applications are data-driven, or trigger oriented. These features coupled with the real-time demands challenged many of the fundamental assumptions for conventional databases and established continuous queries processing on relational data streams as a major research area.  

\subsection{Applications to High-Performance Graph Analytics}
The emergence of continuous subgraph matching on massive data streams can be viewed as a resurrection of the above research theme, albeit in the graph domain.  The paramount importance of timeliness of  detection of a match dictates that the runtime of a query should be dependent on size of the latest information update and independent of the size of the database.  Given the massive scale of the problem and the high computational complexity of the algorithm one is naturally tempted to turn towards established parallel and distributed computing techniques.  Unfortunately, the tools and techniques that perform exceptionally well for traditional HPC applications are not equally effective for graph-theoretic applications due to issues such as irregular memory access and data intensive nature characterized by high data access to computation ratio \cite{DBLP:journals/ppl/LumsdaineGHB07}.  Hence for Area II, we are actively developing an  incremental query processing processing framework \cite{Choudhury:2011:PNNL_Tech_Rep_DAMNET} for massively multi-threaded architectures such as the Cray XMT \cite{Mizell:2009:EEL:1586640.1587769}.  We believe the availability of rich semantic information in terms of labels, timestamps and other attributes in the graph data primes them for leveraging on findings from earlier CQ research.  In the broader context, we are interested in exploring useful but restricted classes of queries where the subgraph matching problem may be tractable (Area I) and will benefit other large-scale graph processing frameworks in an architecture agnostic way.

The contribution of this paper is to draw the attention of the parallel graph mining community to the approaches undertaken by the database community for challenges that are similar in spirit and stimulate discussions on their extension to the graph based applications.  Section \ref{sec:Continuous Query Systems - A Review} presents a brief overview of the continuous query systems research in the database community as well as of research on graph queries.  We present an example graph query and illustrate the associated challenges in section \ref{sec:Continuous Subgraph Matching Framework}.  We review the approaches adapted by the database community to address similar challenges and discuss their relevance for the running example in section \ref{sec:Cross-Cutting Themes}.  Section \ref{sec:Conclusions and Research Plan} concludes with our discussions and suggestions for future research.

\begin{figure}[!t]
\centering
\includegraphics[scale=0.6]{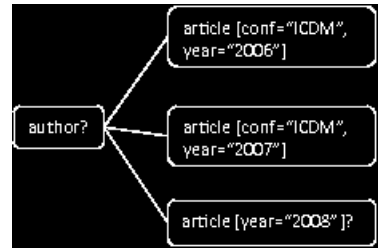}
\caption{Example graph query that emerges over time}
\label{fig:query_graph}
\end{figure}

\begin{figure*}[]
\includegraphics[scale=0.6]{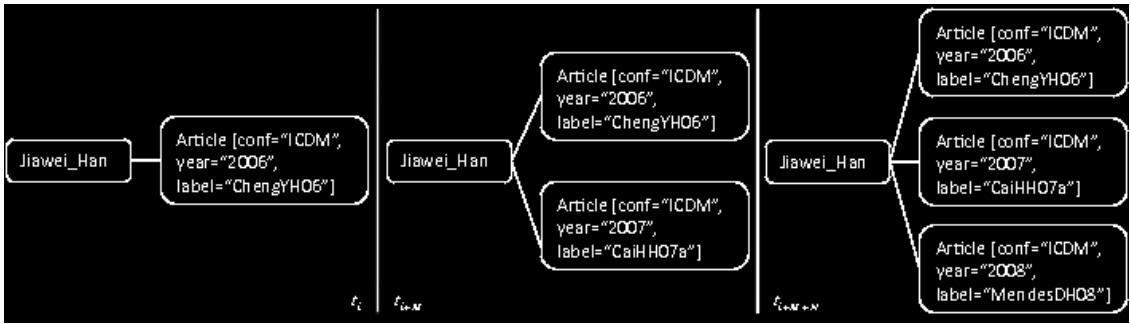}
\caption{Evolution of a matching subgraph}
\label{fig:dynamic_graph}
\end{figure*}

\section {Related Work}
\label{sec:Continuous Query Systems - A Review}
The following sections provide a quick overview of past research from both the database and graph mining community.  For the former, we  omit important topics such as load-shedding on streams or approximate techniques for stream-based queries and restrict our attention to works relevant to exact subgraph matching.
\subsection{Continuous Queries in Databases}
The literature on database research from past the two decades is abundant with work on continuous query systems \cite{Chandrasekaran:2003:TCD:872757.872857, DBLP:conf/sigmod/ArasuBBDIRW03}. A common theme across various solutions are the implementations of window-based operators that exposed a data stream as a buffer or queue on which traditional relational operators as selection, projection and joins were applied \cite{Abadi:2003:ANM:950481.950485}.  A large body of work was dedicated to extending SQL and incorporation of continuous query semantics by adding various constraints \cite{Arasu:2006:CCQ:1146461.1146463,Jain:2008:TSS:1454159.1454179}.  Continuous query processing involved two types of activities like conventional databases.  Query registration involved generating an optimal query plan at compile-time and the run-time processing can be viewed as a dataflow system where a tuple flows through a directed graph of processing operators \cite{Babu:2004:EKC:1016028.1016032}.  However, adaptive query processing systems \cite{Madden:2002:CAC:564691.564698} are an exception to this approach due their dynamic routing of tuples through various operators.  

\subsection{Graph Querying Systems}
Subgraph matching in static graphs has been studied extensively.  Recently, there has been a surge of interest in the database community that seeks to develop hybrid approaches by combining indexing techniques that exploit the distribution of labels in the graph along with graph edit distance or motif based approaches \cite{zhang2010sapper, Zhao:2010:GQO:1920841.1920887, Khan:2011:NBF:1989323.1989418}.  However, investigation of subgraph matching in the context of dynamic graphs did not receive much attention in the broader research community until recently.   The work by Fan et al. \cite{Fan:2010:GPM:1920841.1920878} presents incremental algorithms for graph simulation, bounded simulation and subgraph isomorphism.  They identify a class of queries for the first case where the problem is bounded and optimal.   Our work \cite{Choudhury:2011:PNNL_Tech_Rep_DAMNET} can be viewed as following the same spirit as focusing on a specific class of queries (temporal in this case) and exploring query constraints that would make the problem tractable.  

\section{Incremental Framework for Continuous Subgraph Matching}
\label{sec:Continuous Subgraph Matching Framework}
Consider the example query in Figure \ref{fig:query_graph}.  Suppose we are interested in authors who have published a paper in the past two ICDM conferences and wish to follow their current work. Also assume that the publications database receives a batch of updates after every conference and we seek to query the database after every update for a match.  Figure \ref{fig:dynamic_graph} shows the evolution of a matching subgraph in the publications data.  It is obvious that a complete match will not be found until time instance $(i+M+N)$ but more importantly, an approach employing an algorithm for subgraph matching in static graphs will search the entire graph beginning from scratch and perform a non-significant amount of work every time before reporting no entries matching the query pattern.  An alternate approach is to track partial matches as they appear and augment them as more information becomes available.  

Herein lies the first challenge.  Most prominent algorithms for pattern matching in static graphs are designed to seek complete matches with the pattern and try to aggressively prune their search space using semantic and structural properties of the query graph.  As an example, if searching the database at time $i$ (Figure \ref{fig:dynamic_graph}), a traditional algorithm will begin from a vertex of type author and require that it has at least three neighboring vertices with matching attributes (as shown in Figure \ref{fig:query_graph}).  This check will fail as the graph at time $i$ does not meet this requirement.  On the other hand, a simple incremental algorithm expands its search space by matching an edge at a time.   When the first edge arrives in the graph that matches with any edge in the query graph, a partial match is created with appropriate mapping information.  When the next matching edge arrives, another new partial match is created and the previous partial match is augmented with an additional edge.  Such partial match augmentations will continue until a complete match is found.  While the discussion of complex boundary conditions are beyond the scope of this paper, it is not hard to see the exponential, unbounded growth in the number of partial matches over time with this strategy.

\section {Cross-Cutting Themes}
\label{sec:Cross-Cutting Themes}
\subsection{Temporal Query Constraints}

Apparently, there are a number of options one can adapt to improve the performance.  If an author's first ICDM publication was in 2007, the above approach will create a partial match when that information arrives.  Assuming no out-of-order arrival, this partial match can never yield a complete match in future because our query also requires a matching publication from 2006.  Also, all partial matches which represent an author with one or more 2006 ICDM publications but with no 2007 ICDM publication can be pruned as well.  In short, these advocate strategies which are \textit{time-aware}.  These are precisely the issues discussed in \cite{Babu:2004:EKC:1016028.1016032} and presented as \textit{ordered-arrival constraints}.   However, introduction of an order or exploiting such properties introduces a new set of challenges such as out of order arrival of data.  For real applications the query processor may need to be robust to arrival patterns.  At this point we are actively working on finding a set of representative temporal queries and studying the performance benefits of such constraints.
\textit{Clustered-arrival constraint} is another useful concept in addition to arrival order.  It can be used to describe the acceptable interval between the arrival of tuples in a database or edges in a graph.  The interval can be specified in terms of time or the number of entities/updates.  Implementation of this constraint will require a query processor to prune its set of partial results by removing stale entries.  It can be viewed as a broader version of a windowed query which requires satisfaction of a predicate over the subset of data arriving within the specified time window.  

\subsection {Adaptive Query Processing}
Another important aspect of the research on query constraints involved monitoring the data stream itself for realistic estimates of the constraint parameters \cite{Babu:2004:EKC:1016028.1016032, Madden:2002:CAC:564691.564698}.  In the database context this is accomplished by maintaining synopsis data structures and algorithms that impose minimum overhead.  The objective of this monitoring is twofold.  The fundamental assumption is that once a query is registered and runs over a long period of time, the characteristics of the data stream can change over time.  This fact may stay unknown to the user running the query.  For an example, let's assume a query that monitors packets flowing through multiple networks.   Because the data packets can arrive out of order, a domain driven approach will account for latencies in the network infrastructure as part of the monitoring process.  Further assume that partial matches that do not yield a complete match within a specified time window are excluded from tracking.  If the expected value of these latency intervals drop over time then the query processor will be overestimating that period and maintain a larger set of partial results.  On the other hand, if the expected interval span increased over time then the query processor will be producing potentially incorrect results.  In the graph stream processing context, our foremost objective is to develop strategies for pruning the set of partial matches or being selective about initiating a partial match.  There are several metrics for structural or semantic summaries of a graph dataset that are amenable to scalable implementation. The above mentioned works motivate us to expose those properties to a query optimizer or composer module in a generic fashion.  

\subsection{Multiple Query Processing}
For the example describe above, a different strategy may focus on reducing the number of partial matches maintained in memory.  The number of matches drop dramatically as publications from longer sequence of years are considered; hence, it may decide to check for a specific subset of the match to arrive before spawning a new partial match.  This involves performing a subgraph isomorphism check and can be expensive for even a small substructure.   However, if multiple queries share the same substructure, the cost of the detecting the subset can be compensated.   A real-life query processing system is expected to have a large number of queries registered in it and their efficient handling is an important criteria for scalability.  While scaling up or scaling out is a design choice determined by the application architecture, simply adding more computing resources to match the increase in number of queries is not sufficient or optimal.  Exploiting common workload across multiple continuous queries is a recurrent theme both in the database \cite{Madden:2002:CAC:564691.564698, Munagala:2007:OCQ:1265530.1265561} and graph mining \cite{zhang2010sapper}.

\section {Conclusions and Research Plan}
\label{sec:Conclusions and Research Plan}
The literature on continuous query processing for relational streams is a rich source of guidance for the emerging area of subgraph pattern matching on streams.  As various parallel and distributed computing systems emerge for processing massive scale graphs,  they need to be complemented with appropriate algorithmic approaches for enabling the same innovations accomplished with high throughput relational data stream systems.  This work serves as a reminder that continuous query processing systems are fundamentally different from conventional query processing systems that follow a command-response model and hence, motivates the rethinking of query processing and optimization strategies.  Algorithmic approaches for query processing are obviously different between relational database systems and graph databases;  however, the concepts such as ordered-arrival and other temporal constraints are immediately relevant to their graph querying counterparts.  Areas as adaptive query processing and multiple query processing can be considered as next-level optimization goals once the first-generation of graph based continuous query systems are developed.

\bibliographystyle{abbrv}
\bibliography{../citations_sutanay}
\end{document}